\documentclass[aps,pra,groupedaddress,nofootinbib,notitlepage,showpacs,floatfix,twocolumn]{revtex4-1}

\usepackage{graphicx,graphics,epsfig,subfigure,times,bm,amssymb,amsmath,amsfonts,mathrsfs}
\usepackage[pdfstartview=FitH]{hyperref}
\usepackage{hypernat}
\usepackage{graphicx}
\usepackage[normalem]{ulem}
\usepackage{cleveref}

\newcommand{\beq}{\begin{equation}}
\newcommand{\eneq}{\end{equation}}
\newcommand{\beqnn}{\begin{equation*}}
\newcommand{\eneqnn}{\end{equation*}}
\newcommand{\beqy}{\begin{eqnarray}}
\newcommand{\eneqy}{\end{eqnarray}}
\newcommand{\beqynn}{\begin{eqnarray*}}
\newcommand{\eneqynn}{\end{eqnarray*}}

\newcommand{\ket}[1]{ | #1\rangle}
\newcommand{\bra}[1]{\langle #1 | }

\newcommand{\ignore}[1]{}

\begin{document}

\author{Silvano Garnerone}
\affiliation
{Institute for Quantum Computing, University of Waterloo, Waterloo, ON N2L 3G1, Canada}

\title{Thermodynamic formalism for dissipative quantum walks}

\begin{abstract}
We consider the dynamical properties of dissipative continuous time quantum walks 
on directed graphs.  
Using a large deviation approach we construct a 
thermodynamic formalism allowing us to  
define a dynamical order parameter, 
and to identify transitions between dynamical regimes. 
For a particular class of dissipative quantum walks 
we propose a new quantum generalization of the  
the classical pagerank vector, used to rank the importance 
of nodes in a directed graph. 
We also provide an example where one can characterize the dynamical 
transition from an {effective} classical random walk to a dissipative 
quantum walk as a thermodynamic crossover between 
distinct {dynamical} regimes.    
\end{abstract}

\maketitle

\section{Introduction}
For a system in thermodynamic equilibrium Statistical Mechanics and Thermodynamics 
provide the formalism to characterize different phases of matter, and the transitions from one phase to another \cite{HillPhysics198701}. 
Quantum fluctuations alone can  give origin, at zero temperature, to quantum phase transitions 
\cite{Sachdev2011} whose 
effects can  experimentally be observed in a  critical region at small but finite temperatures. Away from 
thermodynamic equilibrium {engineered 
open quantum systems provide a more recent source of interest  \cite{Verstraete2009,Diehl2008,Kraus2008}. 
In these latter  settings a controlled environment can be used to drive the system 
to a steady state whose properties depend on system and environment parameters}. 
This approach has 
been used to study dissipative phase transitions 
controlled by the coupling constants in the equation of motion \cite{Diehl2010}, 
and to formulate a dissipative model of quantum computation \cite{Verstraete2009}. 

Most of the research on quantum dissipative systems focuses on the 
static properties of the steady state, 
in this work we adopt a different  {dynamical} perspective, 
where the time parameter plays {a more explicit} role in  characterizing the system. 
This allows us to 
study the occurrence of dynamical crossovers or dynamical phase 
transitions driven by fluctuations in the ensemble of possible 
evolutions.
The model systems that we consider, of relevance both to quantum computation and condensed 
matter,  are
\emph{dissipative} Continuous Time Quantum Walks (CTQWs). 
The standard CTQWs (the unitary analog of 
classical random walks) 
describe the coherent motion of an excitation on the vertices of an undirected 
graph \cite{Aharonov2001,Farhi1998,Childs2002,Farhi2008}. 
They are a powerful tool in quantum computation, providing a useful 
approach to algorithms' design, and 
an alternative framework for universal quantum computation \cite{Childs2009}. 
{More recently} CTQWs have also been used 
to describe transport properties in complex networks 
\cite{Mulken2011,Tsomokos2011}, {while} 
dissipative CTQWs have found applications  
in the context of quantum 
biology\cite{Rebentrost2009,Mohseni2008a,Plenio2008,Caruso2009,Rebentrost2009a,
Giorda2011}, 
and quantum state transfer in superconducting qubits 
\cite{Strauch2009,Strauch2008}. 
  
In this work, following \cite{Whitfield2010}, we will refer to 
dissipative CTQWs as Quantum Stochastic Walks (QSWs). 
Using the quantum jump approach \cite{Plenio1997} we study   
the statistical properties of the ensemble of 
trajectories originated by QSWs. 
The mathematical tool allowing us to 
probe the dynamics of the system and the structure of the space 
of trajectories is known as Large Deviation (LD) theory \cite{Touchette2009}. 
The latter 
finds applications in different areas of statistical mechanics and computer science (see \cite{Touchette2009} 
and references therein). 
Using the LD approach it is possible to construct a formalism, 
for the ensemble of possible evolutions, resembling 
standard thermodynamics.  Since the LD theory overlaps with the 
thermodynamic formalism developed  by Ruelle and others \cite{Ruelle2004} 
in the context 
of chaos theory and dynamical systems, 
in the following we will use the expressions thermodynamic 
formalism and LD approach interchangeably. 

From the application of the thermodynamic formalism to QSWs we obtain 
two {main} results: first, a quantum generalization 
of the pagerank vector \cite{Pagerank,citeulike:796239}, 
a measure used to rank the importance of nodes in a graph; second, 
a tool to study transitions from classical to quantum dissipative dynamical regimes 
from a thermodynamic point of view (see \cite{shikano2011differences} for a different approach). 
In particular, we provide an instance of a simple QSW presenting 
a dynamical crossover between the two regimes.

{The rest of the paper is divided as follows:
in the second section we} define the notion of QSW as introduced in 
\cite{Whitfield2010}; 
in the third section we provide a brief introduction to the LD approach;  
in the fourth section we apply the thermodynamic formalism to describe the 
dynamics of a QSW on a directed graphs; while
the last section is devoted to discussions and conclusions. 
 
\section{Quantum stochastic walks}
QSWs are a class of stochastic processes
introduced in \cite{Whitfield2010} with the purpose of 
generalizing both continuous time quantum walks, and 
classical random walks \cite{Venegas-Andraca2008} into a single framework. A QSW 
is specified by a quantum master equation in Lindblad form \cite{breuer2002theory} 
{describing the evolution of the system's density matrix}
\beq
\dot{\rho}= -i [H,\rho] +\sum_{k=1}^{N_L} L_k\rho L_k^\dagger - \frac{1}{2} \{ L_k^\dagger L_k,\rho \},
\label{eq:qme}
\eneq 
where $[\cdot,\cdot]$ and $\{\cdot,\cdot\}$ stand respectively for the commutator and the anticommutator, and 
the operators $L_k$ are called jump operators.
The dynamics described by \cref{eq:qme} is characterized by the interplay between a coherent part 
associated to $H$, and an incoherent part associated to the 
jump operators $\{L_k\}$. If only the coherent term was present the QSW would correspond to 
a standard CTQW. On the other hand, it can be shown that if only the incoherent term was present 
the QSW would correspond to a classical random walk for the diagonal elements of $\rho(t)$  \cite{Whitfield2010}.

A standard stochastic method to study the dynamics described by \cref{eq:qme} is provided by the quantum jump approach \cite{Plenio1997,breuer2002theory}. 
In this framework one considers an ensemble of 
piecewise deterministic evolutions of pure states $\{\ket{\psi_{\alpha}(t)}\}$  
({where} $\alpha$ labels an element in the ensemble). 
The average over different trajectories (or histories) provides 
the state of the system at each instant of time 
\beq
\rho(t) = \left[ \ket{\psi_{\alpha}(t)}\bra{\psi_{\alpha}(t)} \right]_{\rm ave}
\equiv \sum_{\alpha=1}^{N_\alpha}\frac{\ket{\psi_{\alpha}(t)}\bra{\psi_{\alpha}(t)}}{N_\alpha} .
\eneq 
The dynamics of each pure state evolves 
deterministically for some time-interval
according to 
the following non-hermitian effective Hamiltonian
\beq
\label{eq:eff}
H_{\rm eff} \equiv H - i \sum_{k=1}^{N_L} \frac{1}{2} L_{k}^\dagger L_k.
\eneq 
The deterministic evolution is interrupted 
at certain random times $\tau$ by a quantum jump operator $L_k$, which 
projects the state according to 
\beq
\ket{\psi(\tau)}\rightarrow \frac{L_k\ket{\psi(\tau)}}{\Vert L_k\ket{\psi(\tau)} \Vert_2}.
\label{eq:qjump}
\eneq 
In between random jumps the state evolves according to $H_{\rm eff}$.
The use of an ensemble of quantum trajectories 
associated to the QSW in \cref{eq:qme} is also referred to as 
the stochastic unraveling of the quantum master equation.  
More details on this method can be found in \cite{Plenio1997,breuer2002theory}. 

Since the quantum walker evolves on a graph 
we also need to specify how the graph's structure enters into \cref{eq:qme}.
A natural way to proceed is 
to use the set of multi-indices $\{(i,j)\}$ 
(where $i$ and $j$ are the nodes of the graph)
to label the index $k$ in $L_k$
\beq
L_k \rightarrow L_{ij} \equiv M_{ij}\ket{i}\bra{j},
\label{eq:jump}
\eneq
and the matrix $M$ contains the rate coefficients of the jumps from 
$\ket{j}$ to $\ket{i}$, in the site basis. This way the quantum jump 
in \cref{eq:qjump} corresponds to hopping from one site to another  
at a rate given by $M$. 
For the coherent part of the evolution the Hamiltonian $H$ in \cref{eq:qme} 
is identified with the adjacency matrix of the graph with all 
directions removed, so that $H$ is a symmetric matrix. 
Note that only the dissipative part in the master equation is the one that  
encodes for the direction of the edges.
In the next section we introduce the thermodynamic formalism used to study the evolution of a QSW. 

\section{Large deviation theory and thermodynamic formalism}
The dynamics of a stochastic system provides
many useful information that cannot be accessed 
from the knowledge of the stationary state alone.
The large deviation approach is the mathematical framework allowing us to access this extra 
information  \cite{Touchette2009}. In fact, recently, 
its application to classical and quantum 
nonequilibrium dynamics has being crucial to show the presence of dynamical phase 
transitions and dynamical crossover phenomena \cite{Hedges2009, 
Garrahan2010, Garrahan2007}. 
In particular, as first observed in \cite{Garrahan2010}, the quantum jump 
approach provides a quantum stochastic process that is well suited to be analysed 
with the LD theory. In what follows we show how to adapt the method 
developed in \cite{Garrahan2010} to the context of QSWs.

Given a quantum stochastic walker evolving according to \cref{eq:qme}, we are interested in the time record of  
jump events associated to one or more operators $L_k$. In the quantum jump approach
this record is one particular realization of a quantum trajectory. 
Introducing a variable $K$, which counts the number of 
events (or jumps) up to a certain time for a particular jump operator, 
we define the probability
\beq
p(K,t) \equiv {\rm Tr}[\rho(K,t)],
\eneq 
corresponding to the observation of a number $K$ of events up to time $t$.  
$\rho(K,t)$  is 
the density matrix of the system conditioned on the observation of those $K$ events.  
The probability generating function of the distribution $p(K,t)$ is given by 
\beq
Z(s,t) \equiv \sum_{K=0}^{\infty} e^{-s K} p(K,t)= {\rm Tr}\sum_{K=0}^{\infty} e^{-sK} \rho(K,t).
\label{eq:gen}
\eneq 
{We define the auxiliary operator $\rho(s,t)$ as the  Laplace transform of $\rho(K,t)$}
\beq
\rho(s,t) \equiv \sum_{K=0}^{\infty} e^{-sK} \rho(K,t).
\label{eq:rhos}
\eneq 
The long time behavior of the generating function $Z$ has  
the following large deviation form 
\beq
Z(s,t) \approx e^{t\theta(s)}.
\label{eq:ld}
\eneq
The reason why this is the case will be more clear when we consider the dynamics of $\rho(s,t)$ 
{in} the limit 
$t\rightarrow \infty$.
\Cref{eq:ld} defines the function $\theta$ which, in this formalism, 
is the analogue of a thermodynamic free-energy density, while $s$
is interpreted as an intensive thermodynamic variable (the analogue of a chemical potential for example), 
conjugate to the counting variable $K$, and the time parameter plays the role of a volume. 
The function $\theta$ is the fundamental object 
characterizing the dynamics of the process in terms of thermodynamic properties 
in the space of stochastic trajectories (the ensemble of possible time records 
for the counting variable $K$). 
Note that it is possible to consider not just one, but instead a set of jump processes associated to the 
jump operators $\{L_{k_i}\}$, 
and correspondingly a set of counting variables $\{K_i\}$  
can be taken into account. 
Accordingly one  {can} introduce different  intensive parameters $\{s_i\}$, conjugated to $\{K_i\}$. 
Since in the following we  associate one counting variable to each node of a graph, 
we use the vector notation for the variables $\vec{s}$ and $\vec{K}$. 

The dynamical free-energy $\theta$ can be obtained from the 
diagonalization of a generalized Liouville superoperator 
describing the dynamics of the auxiliary operator $\rho(\vec{s},t)$.  
As shown in \cref{sec:A1} the Laplace transform 
in \cref{eq:rhos} decouples the evolution of $\rho(\vec{K},t)$ for different vectors $\vec{K}$, 
and this allows us to derive the following generalized quantum master equation for the 
auxiliary operator 
\beq
\label{eq:gqme}
\dot{\rho}(\vec{s},t)=\mathbb{W}_{\vec{s}} [\rho(\vec{s},t)] \equiv \mathbb{L}[\rho({\vec{s}},t)]+\mathbb{V}_{\vec{s}} [\rho(\vec{s},t)].
\eneq 
$\mathbb{L}$ is the same Liouville superoperator as in \cref{eq:qme} 
\beq
\mathbb{L}[\cdot] =  -i [H,\cdot] +\sum_{k=1}^{N_L} L_k\cdot L_{k}^\dagger - \frac{1}{2} \{ L_{k}^\dagger L_k,\cdot \},
\eneq
while 
\beq
\label{eq:V_s}
\mathbb{V}_{\vec{s}} [\cdot] =\sum_{d} (e^{-s_{d}} - 1)  L_{d} \cdot L_{d}^\dagger,
\eneq
and $d$ labels elements of a subset of $\{1,...,N_L\}$, those associated to the jump processes that we want to track in time. 
When all the entries of $\vec{s}$ are 0 we recover the real typical dynamics, 
otherwise $\mathbb{W}_{\vec{s}}$ determines an 
evolution whose unfolding into trajectories is biased by $e^{-\vec{s}}$. Following \cite{Garrahan2010} we call this ensemble of 
trajectories the $s$-ensemble. 
In the classical context the analogue of $\mathbb{W}_{\vec{s}}$ is known as the Lebowitz-Spohn operator \cite{Lebowitz1999}, 
and the free-energy function $\theta$ is given by the largest real part of its eigenvalues. 
In the quantum case one can adopt a similar prescription and identify $\theta({\vec{s}})$ with the largest real part of the eigenvalues 
of $\mathbb{W}_{\vec{s}}$ \cite{Garrahan2010}, accordingly with \cref{eq:ld} and with 
the fact that for long times the largest real part of the eigenvalues will be the dominant one in evaluating $Z(s,t)$. 
Once we have associated a 
dynamical free-energy to the ensemble of quantum trajectories of the system 
we can consider derivatives of the free-energy function, whose behavior 
tell us about the presence of dynamical crossovers or 
dynamical phase transitions. 
As an example calculation we now derive the form of the first and second derivatives of the dynamic free-energy. 
Using \cref{eq:gen,eq:ld,eq:rhos} we can write in the long time limit
\beq
\theta({\vec{s}})=\frac{d}{dt} {\rm Log} \; {\rm Tr} \rho({\vec{s}},t)=\frac{{\rm Tr}\dot{\rho}({\vec{s}},t)}{Z({\vec{s}},t)}.
\eneq
The numerator in the last expression equals to 
\beq
{\rm Tr}\dot{\rho}({\vec{s}},t)={\rm Tr}\mathbb{W}_{\vec{s}} [{\rho}({\vec{s}},t)]={\rm Tr}\mathbb{V}_{\vec{s}} [{\rho}({\vec{s}},t)],
\eneq
where the last equality follows simply from \cref{eq:qme} and the cyclicity property of the trace.
Hence we have that
\beq
\theta({\vec{s}})=\frac{{\rm Tr}\mathbb{V}_{\vec{s}}[\rho(\vec{s},t)] }{Z({\vec{s}},t)}=\sum_{d}(e^{-{{s_{d}}}}-1){\rm Tr}\; L_{d}^\dagger L_{d} \tilde{\rho}({\vec{s}},t),
\eneq
where we have introduced the normalized density matrix
\beq
\tilde{\rho}({\vec{s}},t) \equiv \frac{\rho({\vec{s}},t)}{Z({\vec{s}},t)}.
\eneq
Notice that $\theta(0)=0$, consistently with the existence of a steady state for the dynamics described by $\mathbb{L}$. 
The partial derivative of $\theta$ with respect to $s_{d}$, for a particular $d$, is given by
\beq
\label{eq:der1}
\partial_{d}\theta\equiv 
\frac{\partial \theta}{\partial s_{d}}=(e^{-s_{d}}-1){\rm Tr}(O_{d} \partial_{d}\tilde{\rho}) - 
e^{-s_{d}}{\rm Tr}(O_{d} \tilde{\rho}),
\eneq
where $O_{d}\equiv L_{d}^\dagger L_{d}$.
We now introduce the {\it activity} vector parameter $\vec{\alpha}$, 
given by the inverse of the first derivative of $\theta(\vec{s})$ 
\beq
\alpha_{d}\equiv-\partial_{d}\theta.
\eneq
It is easy to see that, at $s=0$, the activity coincides with ${\rm Tr}[O_{d} {\rho}^{(ss)}]$;   
${\rho}^{(ss)}$ being the stationary state associated with $\mathbb{L}$. This means that, at $s=0$, the dynamical activity $\vec{\alpha}$ coincides with the 
static expectation value of the operator $O_{d}$ in the steady state, so in this case there is a direct correspondence between  
static and dynamic order parameters. Hence it is not necessary to 
perform a dynamical analysis to access the value of the activity at $s=0$. 
Nevertheless higher order derivatives of the free-energy function, and 
the behavior of the system at $s \neq 0$ cannot be determined by static quantities. 
Those are the kind of information that can be accessed only with the LD
approach. 
For a general value of $\vec{s}$ it is easy to show 
that $t\alpha_d=[K_d]_{\rm ave}$, in fact
\beqy
\frac{[K_d]_{\rm ave}}{t} &=& \frac{\sum K_d e^{-\vec{s}\cdot \vec{K}}p(\vec{K},t)}{t Z}\nonumber\\
&=&\frac{-\partial_d{\rm Log}Z}{t}=-\partial_d\theta=\alpha_d.
\eneqy
This explains why $\vec{\alpha}$ is named activity: 
it coincides with the typical number of jump 
events per unit of time, for a particular value of $\vec{s}$.

The second partial derivative of $\theta$ in the $d$ direction is given by
\begin{eqnarray}
\partial^2_{d}\theta&=&e^{-s_{d}}{\rm Tr} (O_{d} \tilde{\rho}) -2 e^{-s_{d}}{\rm Tr}(O_{d}\partial_{d}\tilde{\rho}) \nonumber
\\ &&+ (e^{-s_{d}}-1){\rm Tr}(O_{d}\partial_{d}^2\tilde{\rho}).
\end{eqnarray}
For the second derivative, even at $s=0$, one does not have a correspondence to any purely static observable. 
A useful quantity to look at in order to 
estimate the fluctuations in time of the local  activity 
is provided by the index of dispersion
\beq
\label{eq:delta}
\delta_{d}(s) \equiv -\frac{\partial^2_{d}\theta(s)}{\partial_{d}\theta(s)},
\eneq 
the ratio between the variance and the mean of the counting variable $K$ for a 
particular jump process. 
In the next section we apply the above formalism in the context of dissipative 
continuous time quantum walks.

\section{Quantum stochastic dynamics on directed graphs}
In this section we study the dynamical properties of a QSW on a directed graph 
using the thermodynamic formalism. 
In particular, we consider the QSW that has been  introduced 
in \cite{sanchez2012quantum} with the purpose of defining a 
quantum analogue of the classical notion of pagerank \cite{Pagerank}. 
The latter is a particular graph centrality measure, i.e. an estimate of the 
relative importance of a node in a graph {with respect to the dynamic of a random walker on the graph}. 
Following \cite{sanchez2012quantum} we consider the  quantum 
master equation given by
\beq
\dot{\rho}= -i [H,\rho] +\sum_{k=1}^{N_L} L_{ij}\rho L_{ij}^\dagger - \frac{1}{2} 
\{ L_{ij}^\dagger L_{ij},\rho \},
\label{eq:google}
\eneq
where, in particular, $H$ is the adjacency matrix of the graph obtained by removing the directions on the edges, 
and the jump operators are defined by 
\beq
L_{ij} \equiv \sqrt{G_{ij}}\ket{i}\bra{j}.
\label{eq:gjump}
\eneq
The matrix $G$, called Google matrix, is a stochastic matrix associated to 
an irreducible and aperiodic classical random walk on the graph 
(for a precise definition of the Google matrix see \citep{citeulike:796239}).
As shown in \cite{sanchez2012quantum}, using a theorem by Spohn \cite{Spohn1977} 
it is possible to prove that the dynamics described by \cref{eq:google} 
is relaxing, i.e. that the stationary state is unique. 
The classical random walk described by the stochastic matrix $G$ 
also has the same property, and its stationary state is called 
pagerank, denoted with $\vec{\pi}$. 
Notice that, without the coherent part given by $H$, the dynamics for the diagonal elements of $\rho(t)$ provided by \cref{eq:google} 
is identical to the classical random walk described by the Google matrix, and in particular its stationary 
state coincides with the classical pagerank \cite{sanchez2012quantum}. 
On the other hand, adding the coherent term $H$ modifies the dynamics and 
the stationary state, allowing for a quantum generalization of the pagerank 
vector. We are now going to use the thermodynamic formalism to detail the differences 
between the purely classical and the more general quantum open dynamics. 

We start calculating the value of the activity at $s=0$ for the set of jump operators 
$\{ L_{ij}; \; j=1,...,n \}$, associated to a chosen node $i$. 
Using \cref{eq:der1,eq:gjump} we have
\beq
\alpha_i(0)=\sum_{j}G_{ij}{\rm Tr }[\ket{j}\bra{j} \rho^{(ss)})=\sum_j G_{ij} \rho_{jj}^{(ss)}.
\eneq
This equation tells us that the activity at $s=0$ is related in a simple 
way to the diagonal entries of the steady state, through the stochastic 
matrix $G$. 
Without the presence of the Hamiltonian $H$ in \cref{eq:google} we would have that $\rho^{(ss)}$ 
is equal to ${\rm diag}(\pi_1,\dots,\pi_n)$, hence
\beq
\alpha_i(0)=\sum_j G_{ij} \rho_{jj}^{(ss)}=\sum_j G_{ij} \pi_{j}=\pi_i,
\eneq
where the last equality follows from the definition of the pagerank vector, the stationary state of $G$. 
This result is simply a check of the consistency of the thermodynamic formalism applied 
to the classical pagerank, and it states that at long times the probability of finding the walker in the   $i$-th node is equal to the average number 
of times the walker hits the $i$-th node. It is interesting to observe that the quantum case is different. When the 
coherent term $H$ is  present the stationary state is different from the mixed classical state associated to the pagerank vector: 
$\rho^{(ss)}\neq{\rm diag}(\pi_1,\dots,\pi_n)$.  Hence the equation
\beq
\label{eq:arank}
\alpha_i(0)=\sum_j G_{ij} \rho_{jj}^{(ss)}
\eneq
defines a \emph{dynamical} quantum analogue of the classical pagerank given by $\vec{\alpha}$, which is 
different from the \emph{static} quantum pagerank $\rho^{(ss)}$ of \cite{sanchez2012quantum}. 
The values provided by $\vec{\alpha}$ quantify
the typical number of times the walker jumps into a particular node in the long time limit. 
On the other hand $\rho^{(ss)}$ provides the probability of finding the walker 
at a particular node. 
While a purely classical dynamics would not distinguish between the 
two quantities, as shown by \cref{eq:arank}, the quantum dissipative 
dynamics does differentiate the two because of the coherences (off-diagonal elements) 
present in the steady state. 

\begin{figure}
\includegraphics[scale=0.3]{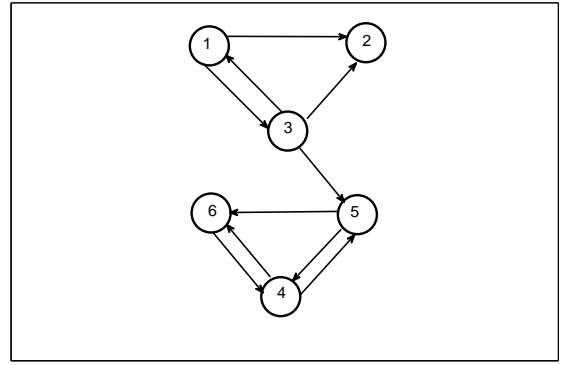} 
\caption{The directed graph that we consider as an example for the 
application of the thermodynamic formalism to QSWs.}
\label{fig:6pag}
\end{figure}

\begin{figure}
\includegraphics[scale=0.3]{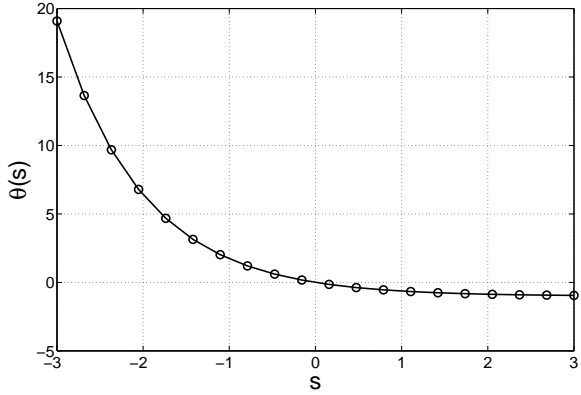} 
\caption{The dynamical free-energy $\theta(s)$ for the QSW associated to the graph in \cref{fig:6pag}.}
\label{fig:theta}
\end{figure}

We consider now, as an example, the QSW on the directed graph shown in 
\cref{fig:6pag}. 
We set the dimension of the vector $\vec{s}$ 
to be equal to the size of the graph and, in order to simplify the discussion,  
we choose all the entries of $\vec{s}$ equal to the same parameter $s$. 
{A similar analysis for a simpler 2-node directed graph can be found in \cref{sec:A2}.}
The dynamical free-energy $\theta$ is plotted in \cref{fig:theta}, 
where one can see that it is equal to 
0 at $s=0$, consistently with the fact that the physical dynamic is relaxing, given that
$\theta$ is the largest real eigenvalue of $\mathbb{W}_s$ in \cref{eq:gqme}.
\begin{figure}
\includegraphics[scale=0.275]{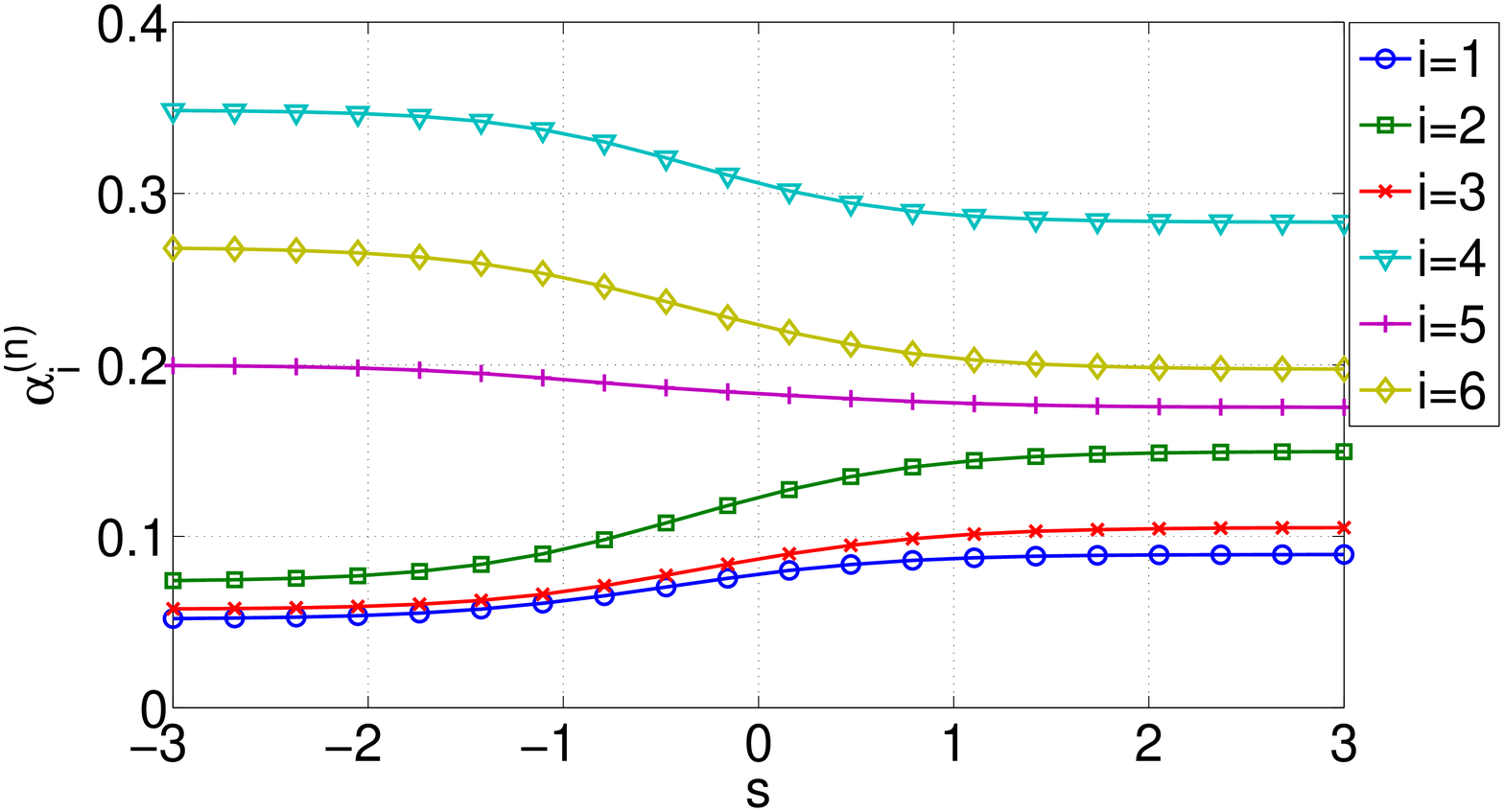} 
\caption{(color online) The normalized dynamical activity of the 
six nodes of the graph in \cref{fig:6pag}. For this particular example 
the ranking provided by the 
activity order parameter does not change as a function of $s$ 
(see text for more details).}
\label{fig:alpha_qtoc}
\end{figure}
\begin{figure}
\includegraphics[scale=0.275]{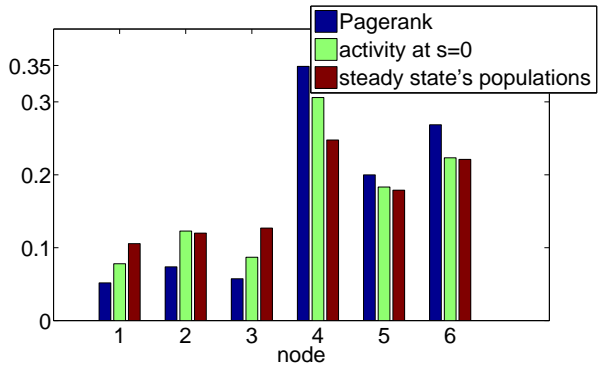} 
\caption{(color online) A comparison between the classical pagerank (left bars), 
the dynamical activity at $s=0$ (middle bars), and the populations in the steady state (right bars). These values are obtained with respect to the graph 
shown in \cref{fig:6pag}. Note that the extreme left values in the plot 
correspond to the classical pagerank.}
\label{fig:comparison}
\end{figure}
More interesting is the behavior of the first two partial derivatives of the free-energy function.  
In \cref{fig:alpha_qtoc} we plot the normalized activity with respect to different sites
\beq
\label{eq:anorm}
\alpha_i^{(n)} \equiv \frac{\alpha_i(\vec{s})}{\sum_{i=1}^6 \alpha_i(\vec{s})}. 
\eneq
The figure shows a smooth dynamical crossover from the active 
region, where large negative values of $s$ favour trajectories with 
many jumps, to the inactive region, where large positive values of $s$ 
favour trajectories with few jumps [see \cref{eq:V_s}]. 
Note that \cref{eq:anorm} is not well defined 
in the limit $s\rightarrow \infty$, since in this  case 
one is left with a purely deterministic evolution according to $H_{\rm eff}$ [see \cref{eq:eff}],  
which implies $\alpha_i(s\rightarrow \infty)=0$. 
This is not a problem for our approach, since one could always 
use the unnormalized activity $\vec{\alpha}$, which is  well defined for every value of $s$. 
However the normalized $\vec{\alpha}^{(n)}$ 
is useful, since it makes the quantitative comparison
between the classical pagerank and its quantum dynamical generalization easier.
Moreover, in the limit 
$s\rightarrow -\infty$ the dynamics in \cref{eq:gqme} is dominated by 
$\mathbb{V}_{\vec{s}}$, with no time intervals of deterministic evolution. 
These two extreme
scenarios select trajectories that are either associated to ``lazy'' QSWs which 
never jumps, or to 
very active QSWs that are effectively equivalent to the  classical 
random walk described by $G$ (in fact $\vec{\alpha}^{(n)}$ in this limit coincides with the 
classical pagerank). 
The transition between the two extreme ensembles of trajectories 
is controlled by the parameter $s$. 
For the particular example that we 
consider the activity ranking (the sorted list of the elements in the vector 
$\vec{\alpha}$)  
is preserved for all values of $s$, so in 
particular it coincides with the classical ranking provided by $\vec{\pi}$, as shown in 
\cref{fig:alpha_qtoc}. 
However the specific values of the activity ranking is a function of $s$, and 
it could be the case that particular graphs show some rank crossings as $s$ is 
changed.
We note that the situation is different 
if one considers the quantum ranking provided by the population (the occupation 
probability) at each 
node in the steady state \cite{sanchez2012quantum}. As  shown in \cref{fig:comparison}  
the population ranking is different for nodes 2 and 3, with respect to the 
{classical} pagerank and the activity ranking. This simple example shows a peculiar aspect of 
QSWs with respect to classical random walks, i.e. their richer dynamical behavior 
translates in distinguishable ways of estimating nodes' centralities.  
\begin{figure}
\includegraphics[scale=0.275]{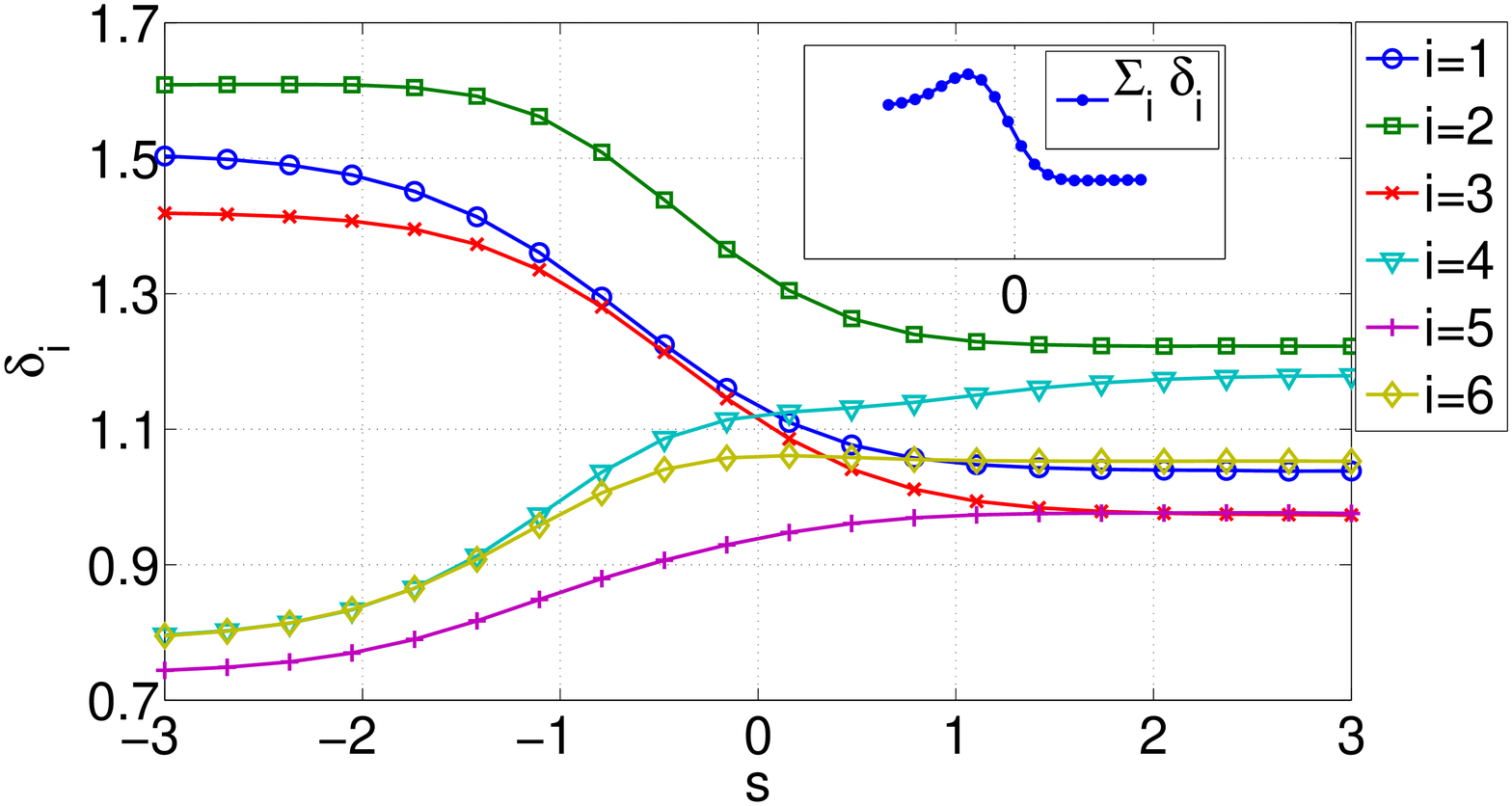} 
\caption{(color online) The stability parameter quantifying  
fluctuations in the trajectories for different values of $s$. 
Each line corresponds to a node in the graph of \cref{fig:6pag}. 
The inset shows the sum with respect to all nodes of the same quantities, 
and it shows a maximum close to {$s=-1$}.}
\label{fig:delta_qtoc}
\end{figure}

\begin{figure}
\includegraphics[scale=0.29]{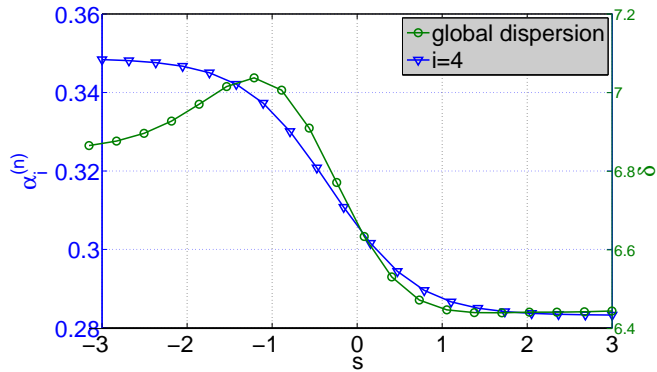} 
\caption{(color online) The figure shows the activity order parameter for the 
4-th node in the graph of \cref{fig:6pag} (triangles), and the total 
fluctuations of the dynamical trajectories on the same graph (circles). 
The presence of a maximum in the fluctuations corresponding to the 
region where the dynamical order parameter is changing is an 
indication of a crossover phenomenon.}
\label{fig:crossover}
\end{figure}

We now look at the dynamical fluctuations of the counting processes. In 
\cref{fig:delta_qtoc} we plot $\delta_i$ [see \cref{eq:delta}]
with respect to different nodes. As we said before 
$\delta_i$ is a measure of the dispersion of the counting variable with respect to 
its average value, 
and it provides information on the fluctuations in time of the stochastic 
process. In 
particular for $\delta_i > 1$ the process is said to be over-dispersed, meaning 
that the jumps are clustered in time, 
having time windows with many jumps, and time windows with almost no jumps. On 
the other 
hand, if $\delta_i < 1$ the process is under-dispersed, meaning that the jump 
events are more 
uniformly distributed in time. The dynamical index of dispersion $\delta_i$ can 
itself 
provide a centrality measure for the $i$-th node, quantifying the stability (or robustness) of 
the activity index 
with respect to fluctuations: the smaller $\delta_i$, the smaller 
are the fluctuations in time of $\alpha_i$. 
{Similarly one can introduce a global index of dispersion: 
$\delta \equiv \sum_i \delta_i$}, 
quantifying the fluctuations at the level of the whole network. 
In \cref{fig:delta_qtoc}  one can see a 
crossover in the dynamical robustness 
from the active region ({associated to an effectively classical dynamics}), 
to the inactive region. Comparing the two extreme behaviours one notices a 
remarkable difference 
in terms of dynamical fluctuations of the local activities. In the classical 
regime  
the top 3 ranked nodes are also the more stable, while the lowest 3 are over-dispersed.
In the quantum regime this is not the case anymore, 
and all the nodes fluctuates with more similar intensities. 
{In general their stability ranking is strongly dependent on $s$  
in a region approximately 
delimited by $s=-2$ and $s=0$}. 
The inset 
in the same figure shows the total index of dispersion, i.e. the sum of all site's indexes, 
which measures the global dynamical fluctuations.  
It is interesting to observe that in the  transition from classical trajectories to 
quantum trajectories the behavior of the total dispersion is not monotonic, 
and it shows a maximum close to {$s=-1$} in the active region. The maximum in the 
global index of dispersion supports the interpretation of a thermodynamic crossover 
from an {effectively} classical to a  quantum regime in the space of trajectories. 
This can be seen more clearly 
in \cref{fig:crossover}, where we plot  the relation between the drop in the 
normalized activity of the 4th sites (the most active), and the global dispersion index. 
From the figure one can identify two distinct values of the activity order 
parameter, corresponding to distinguished dynamical phases, together 
with an increase in dynamical fluctuations in the region 
separating the two, 
{which includes the physically relevant point $s=0$. 
In \cref{fig:delta_qtoc} one can see that the stability ranking is highly dependent on $s$ around $0$, which 
implies that small fluctuations in the dynamical trajectories will have a noticeable effect on $\vec{\delta}_i$}. 

\section{Conclusions and outlook}
In conclusion, we have applied the large deviation approach to study 
the dynamics of quantum stochastic walks, a 
particular case of dissipative quantum evolution on directed graphs. 
The LD approach allows one to formulate 
a thermodynamic formalism describing quantum dissipative evolutions.
We find that the properties of the ensemble of dynamical 
trajectories of the quantum walker can be used to 
characterize the structure of the underlying graph. 
We introduce two new quantum centrality measures \cite{Boccaletti2006}, the 
activity rank and the stability rank.   
These measures are of relevance in the context of complex networks theory
\cite{Barrat2008,barabasi02,Barabasi2005}, providing another 
connection between complex networks and quantum 
information theory  
\cite{Giraud2005,Braunstein2006,Kimble2008,Lapeyre2009,Perseguers2010,Perseguers2010a,Perseguers2010b,
Anand2011,Wu2011,Cuquet2011,PhysRevA.85.062313,Garnerone2012a,2012NatSR...2E.444P,Garnerone2012,sanchez2012quantum,Adhikari2012}.
Another interesting aspect of the thermodynamic approach 
to dissipative quantum walks is the possibility to frame 
dynamical phases of the evolution in the standard language 
of Statistical Mechanics and Thermodynamics. We provide an example 
showing that, already on a small directed graph, the quantum walker 
displays two dynamical regimes separated by a {smooth} thermodynamic crossover. 
{Note that this approach is valid also for undirected graphs, though in this case 
the identification of the most relevant nodes can be simply done looking at the degree distribution: 
the higher the degree of a node, the higher will be its classical or quantum centrality measure; while 
the information related to fluctuations around the typical values is less trivial, and it can be obtained with the 
LD approach.}

It would be of interest to understand the connections 
between the topology of the underlying graph and the dynamical 
phase space of the quantum dissipative process. 
{The comparison between the simple case of a two-node graph provided in \cref{sec:A2} --where no crossover is present-- and the 
more structured graph in \cref{fig:6pag}, suggests that the topology of the underlying graph can be responsible for the occurrence of different 
dynamical phases.}
In particular, exploring more complex graph structures, 
one could characterize the cases where the transition is sharp 
(as found in other physical contexts 
\citep{Garrahan2007,Hedges2009,Garrahan2010}), 
which would imply a dynamical phase transition  
between {effectively} classical and quantum coherent regimes.

\bibliographystyle{apsrev4-1}
\bibliography{biblio}

\appendix
\section{Derivation of the generalized quantum master equation}
\label{sec:A1}
We denote with 
\beq
\vec{K}\equiv [K_1,\dots,K_i,\dots,K_l]
\eneq
the vector corresponding to the occurrence, up to a certain time, 
of a number of $K_i$ events associated to the jump processes described by $L_i$: 
$K_1$ jumps associated to $L_1$, $K_2$ jumps associated to $L_2$ , and so on. 
The probability at time $t$ for the set of events  $\vec{K}$ to occur is given by
\beq
p(\vec{K};t)\equiv {\rm Tr} \rho(\vec{K};t),
\eneq
and $\rho(\vec{K};t)$ is the reduced density matrix of the system conditioned on the 
observation of the set $\vec{K}$ of events. We  define the generating function $Z$ as follows
\beq
Z(\vec{s};t)\equiv \sum_{K_1,\dots,K_l=0}^\infty e^{-\vec{s}\cdot\vec{K}} p(\vec{K};t)=
{\rm Tr} \rho(\vec{s};t),
\eneq
where
\beq
\rho(\vec{s};t)\equiv\sum_{\vec{K}} e^{-\vec{s}\cdot\vec{K}} \rho(\vec{K};t).
\eneq
Using \cref{eq:qme} 
the time evolution of $\rho(\vec{K};t)$, in the $\vec{K}$ sub-manifold, 
is  given by
\beq
\dot{\rho}(\vec{K};t)=\mathbb{L}_0[{\rho}(\vec{K};t)]+\sum_i\mathbb{L}_i[{\rho}(\vec{K}^{(i)};t)],
\eneq
with
\beqy
\mathbb{L}_0[\cdot]&\equiv&-i[H,\cdot]-\frac{1}{2}\sum_i \{L_i^\dagger L_i, \cdot\}, \nonumber\\
\mathbb{L}_i[{\rho}(\vec{K}^{(i)};t)]&\equiv&\sum_i L_i {\rho}(K_1,\dots,K_i-1,\dots,K_l;t) L_i^\dagger.\nonumber\\
\eneqy
It follows that the generalized quantum master equation for $\rho(\vec{s};t)$ is given by
\beqy
&&\dot{\rho}(\vec{s};t)=\sum_{\vec{K}} e^{-\vec{s}\cdot\vec{K}} \dot{\rho}(\vec{K};t)=\nonumber\\
&&\sum_{\vec{K}} e^{-\vec{s}\cdot\vec{K}} \mathbb{L}_0[{\rho}(\vec{K};t)]
+\sum_{\vec{K}} e^{-\vec{s}\cdot\vec{K}} \sum_i\mathbb{L}_i[{\rho}(\vec{K}^{(i)};t)]=\nonumber\\
&&\mathbb{L}_0\left[\sum_{\vec{K}} e^{-\vec{s}\cdot\vec{K}}{\rho}(\vec{K};t)\right]  +
\sum_i  \mathbb{L}_i\left[ \sum_{\vec{K}} e^{-\vec{s}\cdot\vec{K}} {\rho}(\vec{K}^{(i)};t)\right]= \nonumber\\
&&\mathbb{L}_0[{\rho}(\vec{s};t)]+\sum_i e^{-s_i} \nonumber\\
&& \times\mathbb{L}_i\left[\sum_{\{K_{j\neq i}\}=0}^\infty e^{-\sum_{j\neq i}s_j K_j} \sum_{K_i=1}^\infty e^{-s_i(K_i-1)}\rho(\vec{K}^{(i)};t) \right] =\nonumber\\
&&\mathbb{L}_0[{\rho}(\vec{s};t)]+\sum_i e^{-s_i}\nonumber\\
&&\times\mathbb{L}_i\left[\sum_{\{K_{j\neq i}\}=0}^\infty e^{-\sum_{j\neq i}s_j K_j} \sum_{K_i=0}^\infty e^{-s_iK_i}\rho(\vec{K};t) \right]=\nonumber\\
&& \mathbb{L}_0\left[\rho(\vec{s};t)\right]+\sum_ie^{-s_i}\mathbb{L}_i\left[\rho(\vec{s};t)\right]\equiv \mathbb{W}_{\vec{s}}\left[\rho(\vec{s};t)\right].
\eneqy
In the above derivation we simply made use of the linearity of the operators, and in the third equality we wrote the $\sum_{K_i=0}^\infty$ into a second equivalent way. 

\section{Two-node directed graph}
\label{sec:A2}
\begin{figure}
\includegraphics[scale=0.3]{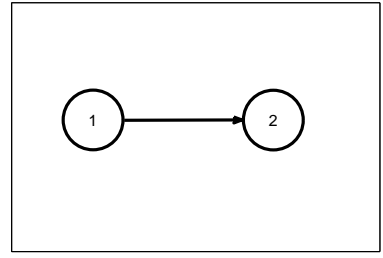} 
\caption{A directed two node graph. }
\label{fig:graph2}
\end{figure}

\begin{figure}
\includegraphics[scale=0.3]{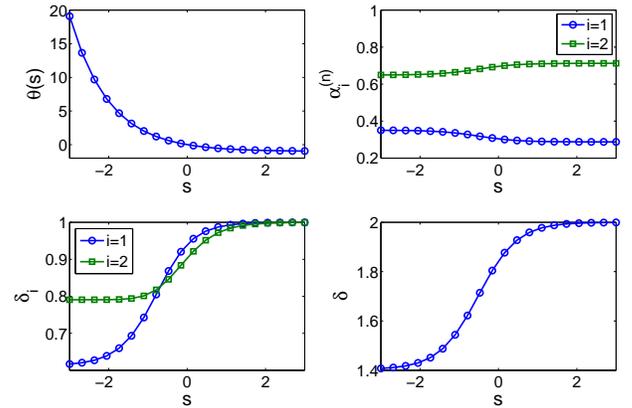} 
\caption{(color online) Simulation results for the directed 2-node graph in \cref{fig:graph2}. See the main text for comments.  }
\label{fig:2site}
\end{figure}

{The most simple example of a directed graph is given by a two-node graph with a directed link  
(see \cref{fig:graph2}). This example is useful because it shows that a trivial graph structure does not show the dynamical 
crossover present in \cref{fig:crossover}. 
\Cref{fig:2site} shows the result of the simulation. The function $\theta$ has the expected monotonic decreasing behavior as a function of $s$. 
The normalized activities $\alpha^{(n)}_{i=1,2}$ of the two nodes   deviate most going from the active to the inactive dynamical region, which is 
different from the case of \cref{fig:6pag}, where more active trajectories allow to better distinguish the activities of the nodes. The index of 
dispersion $\delta_{i=1,2}$ increases in a smooth way for both nodes from the active to the inactive region, approaching a limiting value of 1, 
characteristic of a Poisson process. The global index of dispersion $\delta$ does not show any maximum between the active and the inactive region, 
implying the absence of dynamical crossovers between the two regimes, supporting the conjecture that more interesting dynamical effects (like 
crossovers or phase transitions) are associated to more structured graphs.  
}

\end{document}